\documentclass{elsart}
\usepackage[dvips]{graphicx}
\usepackage{amsfonts}
\usepackage{lscape}
\begin {document}
\begin{frontmatter}
\title{Extraction of information about periodic orbits from scattering
functions}
\author[Basel]{T. B\"{u}tikofer}
\author[CIC]{C. Jung}
\author[CIC]{T.H. Seligman}
\address[Basel]{Institut f\"{u}r Physik, Universit\"{a}t Basel, CH-4056
Basel, Switzerland}
\address[CIC]{Centro de Ciencias F\'\i sicas, University of Mexico
(UNAM), Cuernavaca, Morelos, Mexico}
\begin{abstract}
As a contribution to the inverse scattering problem for classical
chaotic systems, we show that we can select sequences of intervals of
continuity, each of which yields the information about period,
eigenvalue and symmetry of one unstable periodic orbit.
\end{abstract}
\end{frontmatter}

In the framework of the study of the inverse scattering problem for
classical chaotic scattering in two dimensions, the attention has been
focussed on the topology of the chaotic saddle \cite{JLS99,Guadalajara}
and on the definition of an appropriate partition \cite{LiJu99};
thermodynamic quantities were also discussed marginally
\cite{JLS99}.
In none of these approaches the specific properties of the unstable
periodic orbits were explicitly searched for.
Yet knowledge of these, goes a long way towards understanding the
chaotic saddle.
Usually the shortest orbits overshadow the chaotic set or, in other
words, they form the skeleton of the globally unstable component of the
invariant set.
As far as semi-classical approximations are concerned, these orbits form
the backbone of all considerations that lead to trace formulae
\cite{Selb56,BaBl74,Gutzwiller}.
The most useful informations are their periods and Lyapunov exponents.
These can be used to determine the  hierarchical order in the pattern of
intervals of continuity (henceforth abbreviated to IOC) in the
scattering functions, but at the same time it is this very pattern,
which allows us to learn something about the periodic orbits from
scattering functions.

In this article we shall focus on the latter part, i.e. on obtaining the
periods and the Lyapunov exponents of unstable periodic orbits from
regular patterns of the IOC of scattering functions.
The latter we define as a function of the points on a line in the space
of initial conditions, that gives some property of the scattering
process such as e.g. time delay or scattering angle (cf
fig.~\ref{t_delay_062}).
After recalling some known results \cite{JuPo89} concerning the external
periodic orbits we present two methods to achieve our goal for periodic
orbits that are both fairly short and not too unstable.
We illustrate these methods by scattering a charged particle off a
magnetic dipole.
The reader may want to refer to the figures of the example while reading
the description of the general method.

The trajectories belonging to one IOC of a scattering function all leave
the interaction region crossing the same external orbit i.e. leaving
through the same saddle or ``doorway''.
When the initial condition approaches one of the boundary points of the
IOC from the inside, the scattering trajectory executes an increasing
number of revolutions along the saddle orbit before it finally leaves
the interaction region.
For each revolution we expect an oscillation of the scattering angle
(cf fig.~\ref{sequence_062}).
The frequency of this oscillations increases with the diverging time
delay.
An asymptotic observer can determine a sequence of initial conditions
from within one and the same IOC, which are separated by a full
oscillation of the scattering angle.
This sequence converges to the boundary of the IOC in such a way, that,
in the limit, the distance from the boundary point decreases by a factor
determined by the eigenvalue of the saddle orbit.
In the same limit the time delay increases by the period of this orbit,
when going from one member of this sequence to the next.
From these two quantities the Lyapunov exponent can be calculated.
If there are several external orbits we can investigate each of them
separately, choosing an appropriate IOC, where the outgoing scattering
trajectories cross exactly this orbit before finally leaving the
interaction region.
Information about any of the internal orbits cannot be obtained in this
way.

Based on this result we propose two methods to obtain both periods and
Lyapunov exponents for inner periodic orbits.
The first will use the time delay function only whereas the second
exploits in addition the other scattering functions.
To achieve this we analyze how the inner orbits influence the
hierarchical pattern of IOC.

The basic idea for our method is provided by the following property of
chaotic scattering: To each periodic orbit $\sigma$, which can be
reached from the outside, there exist infinitely many sequences of IOC
(cf fig.~\ref{t_delay_062}) of the scattering functions. 
The common feature of these sequences is the following:
As we move along the sequence, sooner or later one step implies
precisely one revolution of the trajectory close to the periodic orbit
$\sigma$.
The central problem will be to identify a sequence which is clearly
associated with one periodic orbit.
If we knew such sequences in advance, then it would be easy to find the
period and the Lyapunov exponent of the periodic orbit $\sigma$.

We label the IOC forming the selected sequence by
$\{I_{\sigma}^{(k)}\}_{k\in\Nset}$ where $k$ labels the
order of the elements of the sequence, $L(I_{\sigma}^{(k)})$
indicates the length of the IOC and $t(I_{\sigma}^{(k)})$ a
typical time delay for this IOC which we obtain by choosing a
representative trajectory for each IOC.
It is convenient to choose the representatives to have the absolute
minimal time delay within an IOC. 

At large time delays two consecutive representatives differ only in one
additional revolution along the unstable periodic orbit.
In order to extract periods and Lyapunov exponents of unstable periodic
orbits from the time delay function, we need to know the representative
time delays and lengths of as many IOC as possible.
We determine the ratio
$\tilde{\lambda}_{\sigma}^{(k)}=L(I_{\sigma}^{(k)})/
L(I_{\sigma}^{(k+1)})$
of the lengths of consecutive pairs of IOC and the difference of
their representative time delays
$\tilde{T}_{\sigma}^{(k)}=t(I_{\sigma}^{(k+1)})-
t(I_{\sigma}^{(k)})$.
We also define
\begin{equation}
\tilde{\Lambda}_{\sigma}^{(k)}=
\frac{1}{\tilde{T}_{\sigma}^{(k)}}
\log{(\tilde{\lambda}_{\sigma}^{(k)})}.
\label{CalcLyap}
\end{equation}
If the sequence was chosen appropriately,
$\tilde{T}_{\sigma}^{(k)}$ will converge to the period
$T_{\sigma}$ of the orbit $\sigma$,
$\tilde{\lambda}_{\sigma}^{(k)}$ to its eigenvalue
$\lambda_{\sigma}$ and
$\tilde{\Lambda}_{\sigma}^{(k)}$ to its Lyapunov exponent
$\Lambda_{\sigma}$.

As mentioned above the choice of the sequence is critical and difficult.
We therefore proceed to show how this choice can be partially avoided
simultaneously with the approximative evaluation of the quantities
above.
If we plot either $\tilde{\lambda}_{\sigma}^{(k)}$ or
$\tilde{\Lambda}_{\sigma}^{(k)}$ against
$\tilde{T}_{\sigma}^{(k)}$ we should find an accumulation point
near the correct value of
($\lambda_{\sigma}$,$T_{\sigma}$) or
($\Lambda_{\sigma}$,$T_{\sigma}$)
respectively.
Note that each of the infinitely many sequences associated
to one periodic orbit will approach the same accumulation point.
Furthermore points
($\tilde{\lambda}_{\sigma'}$,$\tilde{T}_{\sigma'}$)
or ($\tilde{\Lambda}_{\sigma'}$,$\tilde{T}_{\sigma'}$)
erroneously taken because they belong to a sequence for a different
periodic orbit will have different accumulation points, except if both
period and eigenvalue for the two orbits coincide.
If the error is such that the pairs are obtained from IOC of sequences
that belong to different periodic orbits no accumulation points are
expected.
It is thus tempting to perform this plot using points generated for all
pairs of IOC available.

In principle we should then find accumulation points for many periodic
orbits.
This will not be implemented easily because if we have sufficient IOC
to expect reasonable convergence along one sequence, the plot would have
so many points that it would be very difficult to identify these
accumulation points.
Even worse, the discrete numerics may simulate structures that do not
exist.
In order to identify the accumulation points, only pairs of IOC that
might belong to the same sequence have to be taken into account.
Appropriate strategies to achieve this goal must be developed.
An inspection of the hierarchical structure corresponding to the time
delay function shows that successive IOC have to be connected by
pieces of equal or higher hierarchical order.
Therefore only pairs of IOC are tested with no intervening IOC
possessing lower time delays.
Further filtering may well be necessary.
\begin{itemize}
\item 
If we proceed using the time delay function as the only source of
information, we can start to look for multiplets of IOC which apparently
belong to one sequence.
Each of its members should have a higher representative time delay than
its predecessor without intervening IOC with lower time delays.
If such a multiplet belongs to a single sequence its successive
IOC should show roughly the same differences in time delay and
ratios of successive lengths (cf fig.~\ref{analyzed_delay062}).
By narrowing the limits for the allowed differences in time delay and
ratios of the multiplets, improved convergence towards a periodic orbit
can be promoted.
\item
If other scattering functions are also available, pairs of IOC that
belong to a sequence can be identified by comparing their representative
final conditions.
Again only pairs of IOC without intervening IOC with lower
representative time delays are taken into account.
Since the final conditions of representative trajectories converge in
the limit of high time delays to a point, two successive IOC can be
tentatively identified by looking for those with close representative
final conditions.
\end{itemize}
Both methods yield not only the basic periods but also, in a diminished
extent, integer multiples thereof.
\\
\\
To illustrate these methods we use the scattering of a charged point
particle off a magnetic dipole.
This model is also known as the St\"ormer problem \cite{Stormer}.
After eliminating the rotational degree of freedom and a suitable
rescaling in cylindrical coordinates one is left with a Hamiltonian of
the form:
\begin{equation}
H(\rho,z)=\frac{1}{2}\left[p_\rho^2+p_z^2+\left(\frac{1}{\rho}-
\frac{\rho}{\sqrt{\rho^2+z^2}^3}\right)^2\right].
\label{Hamiltonian}
\end{equation}
Here $(p_\rho,p_z)$ are the momenta conjugate to $(\rho,z)$.
It is well known that the resulting phase flow represents a chaotic
scattering system \cite{DrFi76,RuJu94b}.
An analytical proof for its non-integrability is given in \cite{AMY92}.
Every scattering trajectory is uniquely determined in the asymptotic
region by the impact parameter $b_{\mathrm{in}}$ and the angle
$\alpha_{\mathrm{in}}$ between its velocity vector and the negative
$\rho$-axis.
A detailed description of the chaotic dynamics of this system for the
energy range $[0.031,0.081]$ can be found in \cite{RuJu94b}.
The two fundamental periodic orbits of the binary horseshoe will be
called $A$ and $C$.
The outer orbit $C$ is accessible from the outside for all energies
whereas the inner one $A$ may be at the center of an elliptic island
for some energies.
\\
For illustration we will consider the hyperbolic case at an energy
$E=0.062$ and show how to extract information on $A$ and other inner
periodic orbits.
Note that hyperbolicity is no prerequisite for the methods presented
here.
For the integration of the equations of motion a Bulirsch-Stoer
algorithm \cite{NumRec} has been used.
The simulation takes place in a disk of radius $1000$ around the origin
of the $(\rho,z)$-plane.\\
A part of a typical sequence whose trajectories come close to $A$ is
shown as IOC in a time delay function in fig.~\ref{t_delay_062}.
In fig.~\ref{sequence_062} the final angles $\alpha_{\mathrm{out}}$ of
the trajectories from the IOC $\{I_A^{(k)}\}_{1\le k\le 5}$ are
plotted against their corresponding time delay values.

In this plot several essential features can be seen.
\begin{enumerate}
\item As the time delay increases the period of the curves converge
towards the period of the outer periodic orbit $C$.
This is a direct consequence of the additional revolutions near $C$ with
initial conditions near the boundary of an IOC.
\item The differences of the representative (minimal) time delays
$t(I_A^{(k)})$ of consecutive IOC converges towards half the
period of the inner periodic orbit $A$.
\item The lower end of the curves of every second IOC point in the
same direction.
This effect has its origin in the common reflection symmetry on the
$\rho$-axis of $A$ and the
Hamiltonian.
If $A$ and the Hamiltonian had no common symmetry
\begin{equation}
t(I_A^{(k)})=\frac{1}{2}\left(t(I_A^{(k-1)})+t(I_A^{(k+1)})\right)
\end{equation}
would not be fulfilled.
In such a case there would be two sequences whose members were separated
by the full period of the corresponding periodic orbit.
This shows that by using only the time delay function as a source of
information it might be unclear whether we measure the whole period of
a periodic orbit without symmetries or its $n$-th part of an periodic
orbit with a cyclic symmetry group of order $n$.
Using additional information provided by other scattering functions as
$\alpha_{\mathrm{out}}$ and $b_{\mathrm{out}}$ can give helpful hints to
the answer of this question.
\end{enumerate}
In order to resolve about $608$ IOC we scatter $10^7$ trajectories
with $\alpha_{\mathrm{in}}=0^\circ$ and $b_{\mathrm{in}}$ evenly
distributed on the interval [1.3049,1.3065].
If we apply the filter that respects the hierarchical structure and look
for triples of IOC $(I_{\sigma}^{(k-1)},
I_{\sigma}^{(k)}, I_{\sigma}^{(k+1)})$
with the property
\begin{eqnarray}
\left|\tilde{T}^{(k-1)}_{\sigma}
-\tilde{T}^{(k)}_{\sigma}\right|&<&0.02\\
\left|\tilde{\Lambda}^{(k-1)}_{\sigma}
-\tilde{\Lambda}^{(k)}_{\sigma}\right|&<&0.02\nonumber
\end{eqnarray}
we end up with $64$ of such triples.
Each triple of IOC yields two points
$(\tilde{\Lambda}^{(k)},\tilde{T}^{(k)})$ in
fig.~\ref{analyzed_delay062}.
Not all $128$ points are located in the area shown in this figure but
the chosen resolution of $10^7$ points was too poor for a meaningful
prediction of orbits with symmetry reduced periods longer than $T=15$.
The more dots are clustered in a group the likelier it is to find a
periodic orbit with roughly these values for its symmetry reduced period
and Lyapunov exponent.
In fig.~\ref{analyzed_delay062}, apart from the times for one revolution
around a symmetry reduced periodic orbit $(A^{1/2},C^{1/2})$ integer
multiples of these times are also recognizable $(A,C)$.
Among the periodic orbits of our system shown in
fig.~\ref{trajectories_062} only three orbits ($A$,$C$ and $I$) remain
invariant under a reflection on the $\rho$-axis.
The rest of the orbits either change their direction of rotation
($B$,$D$,$F$,$G$ and $H$) or are mapped onto their mirror image ($E$).
Thus for orbits $A$,$C$ and $I$ the half periods show up besides the
full ones in fig.~\ref{analyzed_delay062}.

Alternatively we may use pairs with similar final conditions retaining
the restriction derived from the hierarchical structure of the time
delay function.
We accept pairs of IOC whose representative trajectories differ in
outgoing angle by less than $0.1^{o}$ and in the outgoing impact
parameter by less than $0.01$ in our scale.
We plot the $115$ obtained points in fig.~\ref{analyzed_062} similar
to fig.~\ref{analyzed_delay062}.
Now half periods will not appear because we did not symmetrize the
outgoing asymptotic parameter.
Comparing the two figures we see that the second method allows to
identify more periods.
While in fig.~\ref{analyzed_delay062} accumulation points are not very
obvious in fig.~\ref{analyzed_062} they appear rather clearly in a few
instances.
Where they appear the Lyapunov exponent indeed seems to converge toward
the exact value.
Note the difference in scales for time delay and exponents.
Considering that the Lyapunov exponent is an average, once identified,
the points belonging to one period can be averaged.
Table~\ref{Result_Table062} shows that for the points in
fig.~\ref{analyzed_062} the average matches the exact value quite well.

Summarising we have presented a new approach to the inverse scattering
problem for chaotic Hamiltonian systems.
In distinction to earlier work \cite{JLS99,Guadalajara,LiJu99} we do not
require the system to have an internal or external clock.
Furthermore we are able to extract more detailed information about the
properties of the most important periodic orbits.
The possibilities inherent to this approach are much larger than what
was presented in this letter.
More extensive research on the filtering techniques must be done and the
optimal approach may well depend on the problem.
Also much larger numbers of intervals of continuity must be generated to
be able to analyze the structure of the
$(\tilde{\Lambda},\tilde{T})$-plane.
Finally the possibility to obtain quite easily the periods of the
shortest periodic orbits may be used to provide an inner clock that
always exist for the method proposed in \cite{JLS99}.

\ack
The authors wish to thank L.~Benet for useful discussions.
This work was partially supported by the SNF, the DGAPA (UNAM) project 
IN-102597 and the CONACYT grant 25192-E.
One of the authors (T.B.) wants to thank the CIC for their generous
hospitality.

\bibliographystyle{unsrt}
\bibliography{references}
\setcounter{figure}{0}
\setcounter{table}{0}
\newpage
\begin{landscape}
\begin{figure}
\begin{center}
\includegraphics[height=13cm]{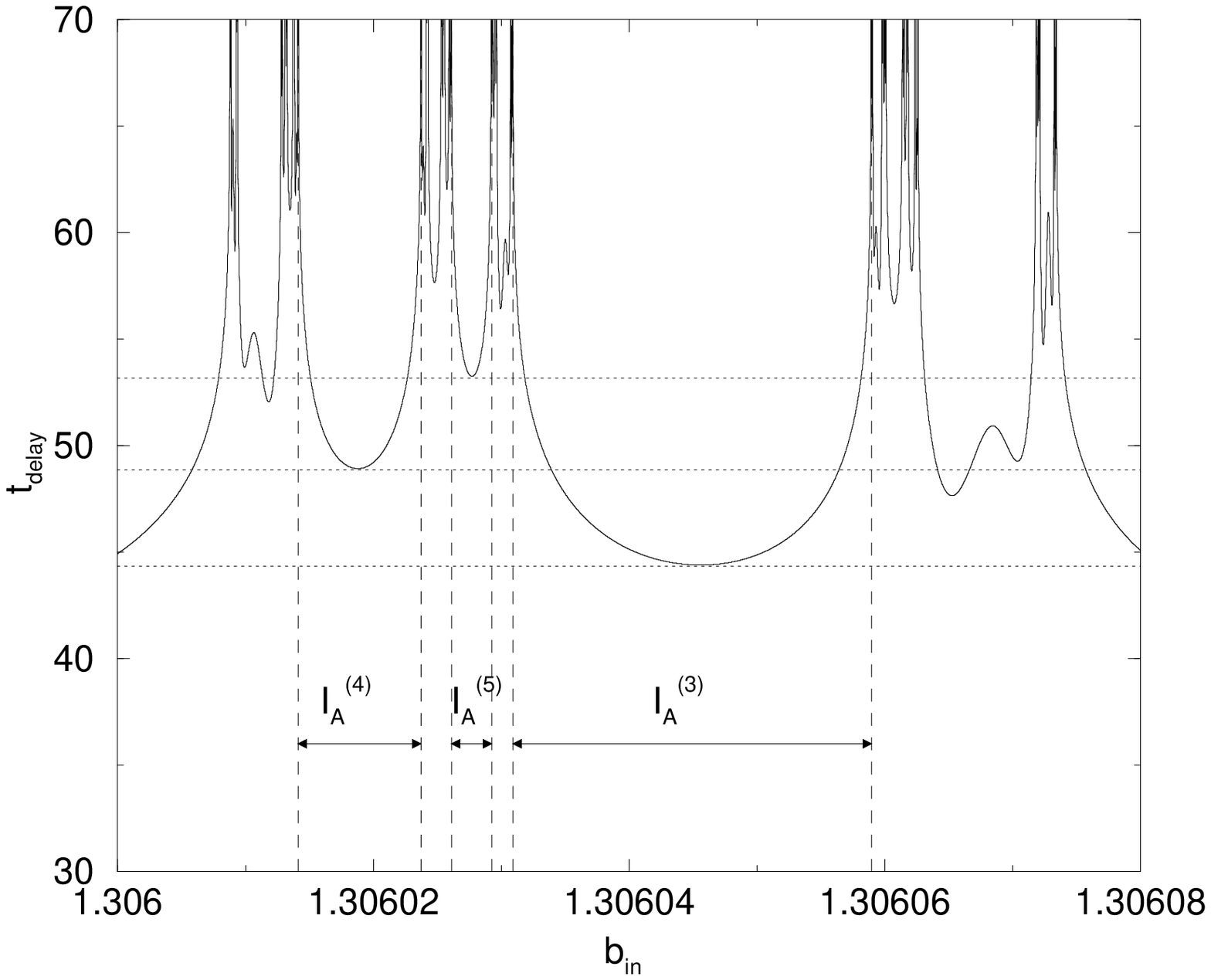}
\caption{Time delay function for $\alpha_{\mathrm{in}}=0$, $E=0.062$.
The consecutive IOC $\{I_A^{(k)}\}_{3\le k\le 5}$ are indicated.}
\label{t_delay_062}
\end{center}
\end{figure}
\end{landscape}
\newpage
\begin{landscape}
\begin{figure}
\begin{center}
\includegraphics[height=13cm]{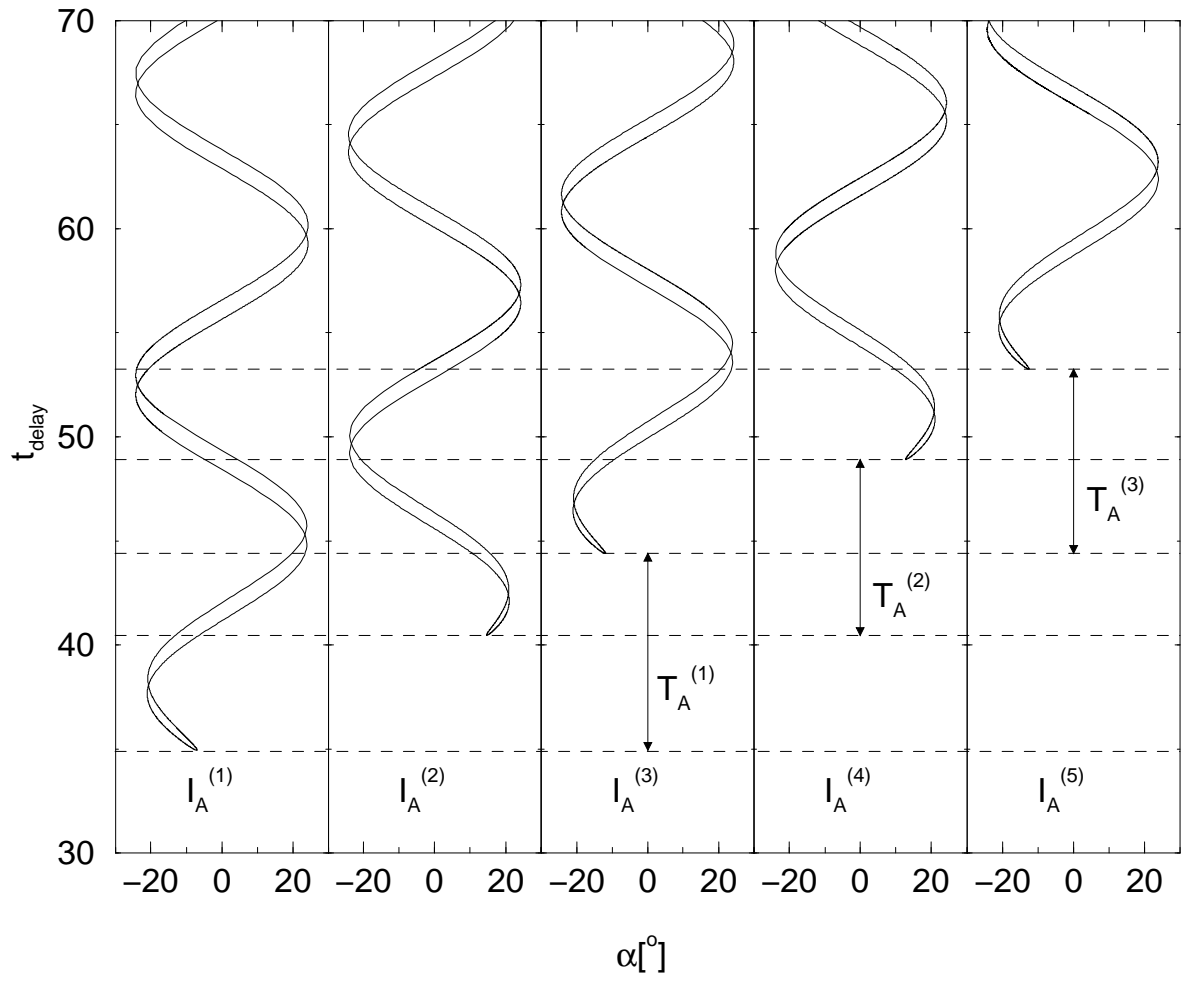}
\caption{$\alpha_{\mathrm{out}}$ plotted against $t_{delay}$ for the IOC
of continuity $I_{A}^{(1)}$,$I_{A}^{(2)}$,$I_{A}^{(3)}$,$I_{A}^{(4)}$,
$I_{A}^{(5)}$.}
\label{sequence_062}
\end{center}
\end{figure}
\end{landscape}
\newpage
\begin{landscape}
\begin{figure}
\begin{center}
\includegraphics[height=13cm]{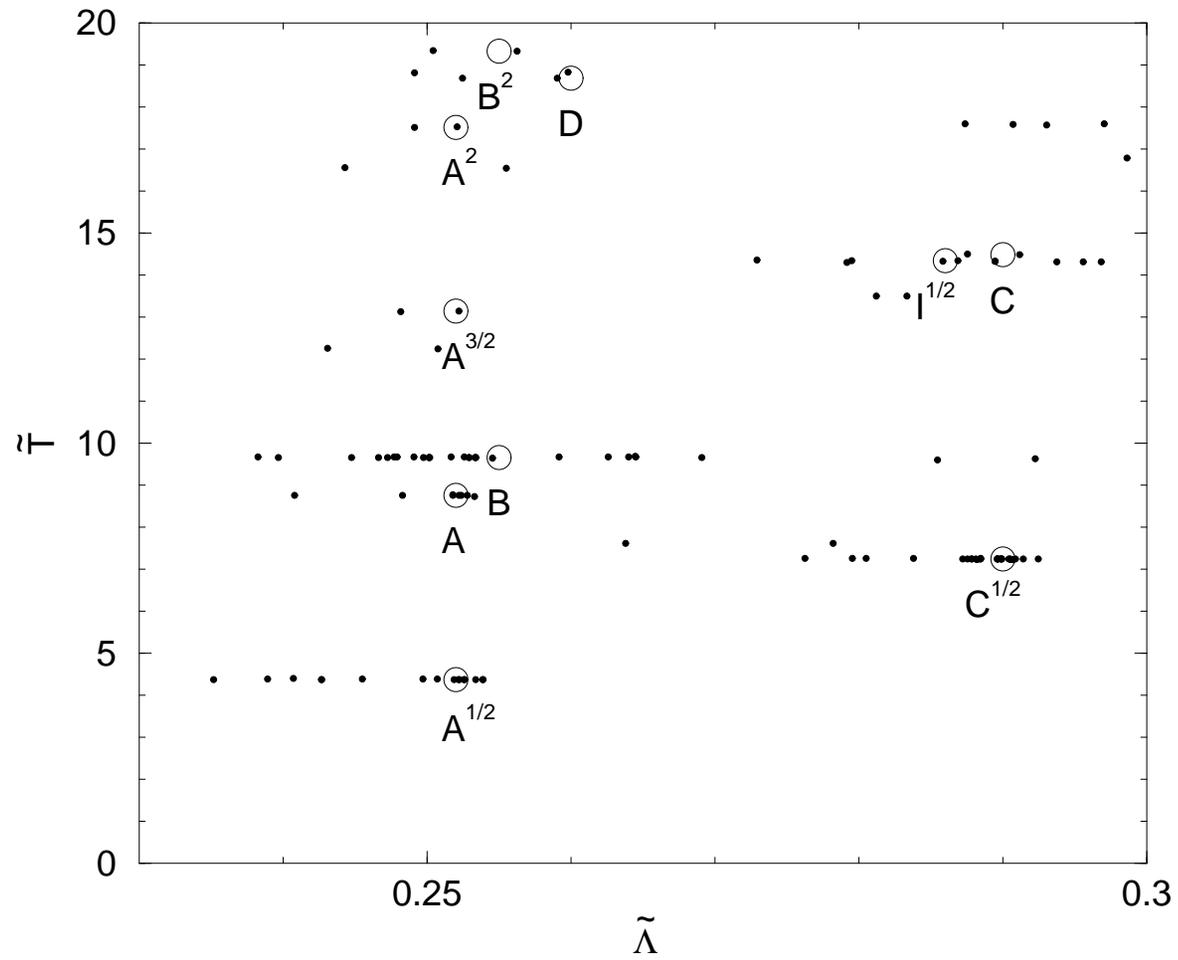}
\caption{Periods and Lyapunov exponents of periodic orbits evaluated by
only using the time delay function as described in the text.
Exact values are marked by open circles and capital letters.
The letter's upper indices display what fraction of the full period
were measured.}
\label{analyzed_delay062}
\end{center}
\end{figure}
\end{landscape}
\newpage
\begin{landscape}
\begin{figure}
\begin{center}
\includegraphics[height=13cm]{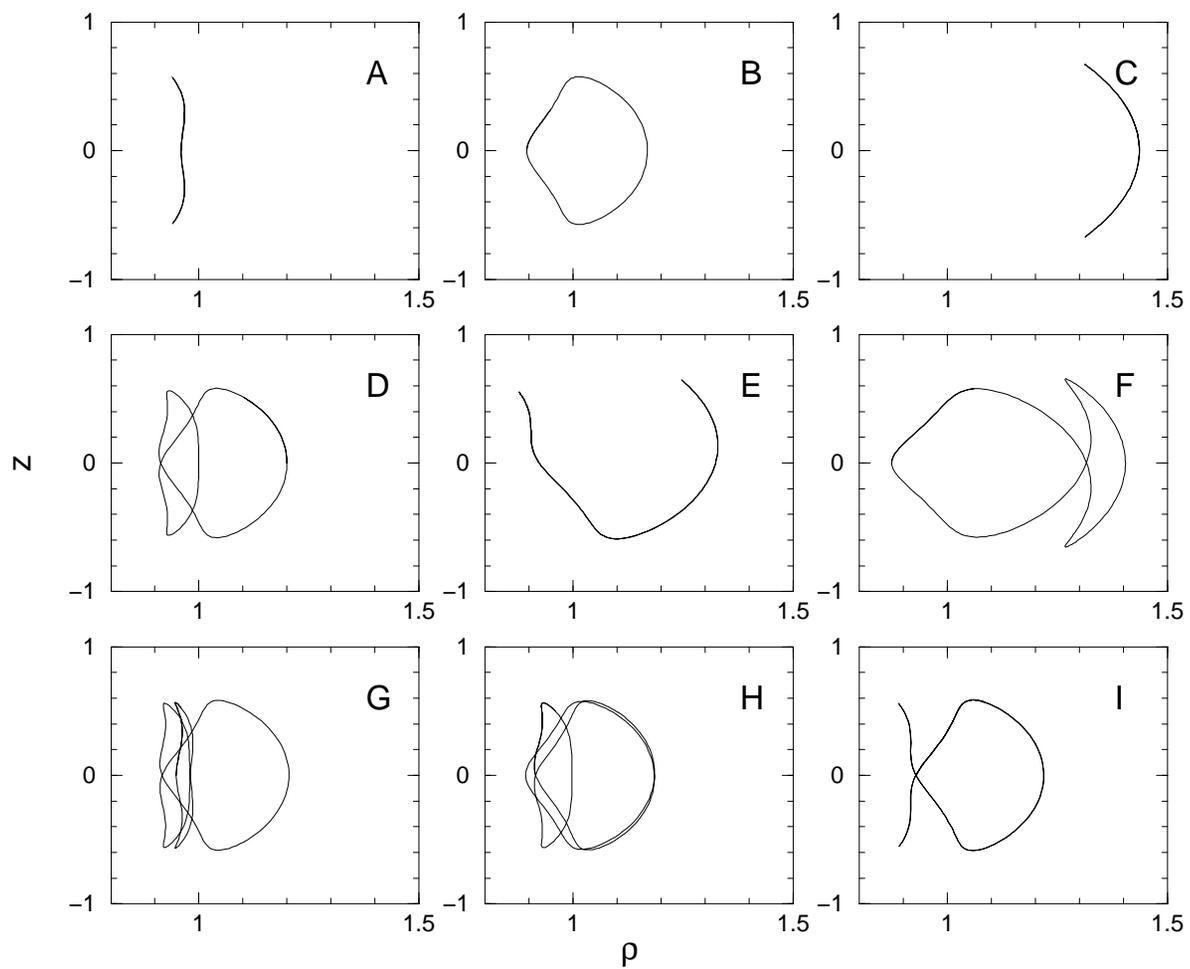}
\caption{Projection of periodic orbits in the $(\rho,z)$-plane.}
\label{trajectories_062}
\end{center}
\end{figure}
\end{landscape}
\newpage
\begin{landscape}
\begin{figure}
\begin{center}
\includegraphics[height=13cm]{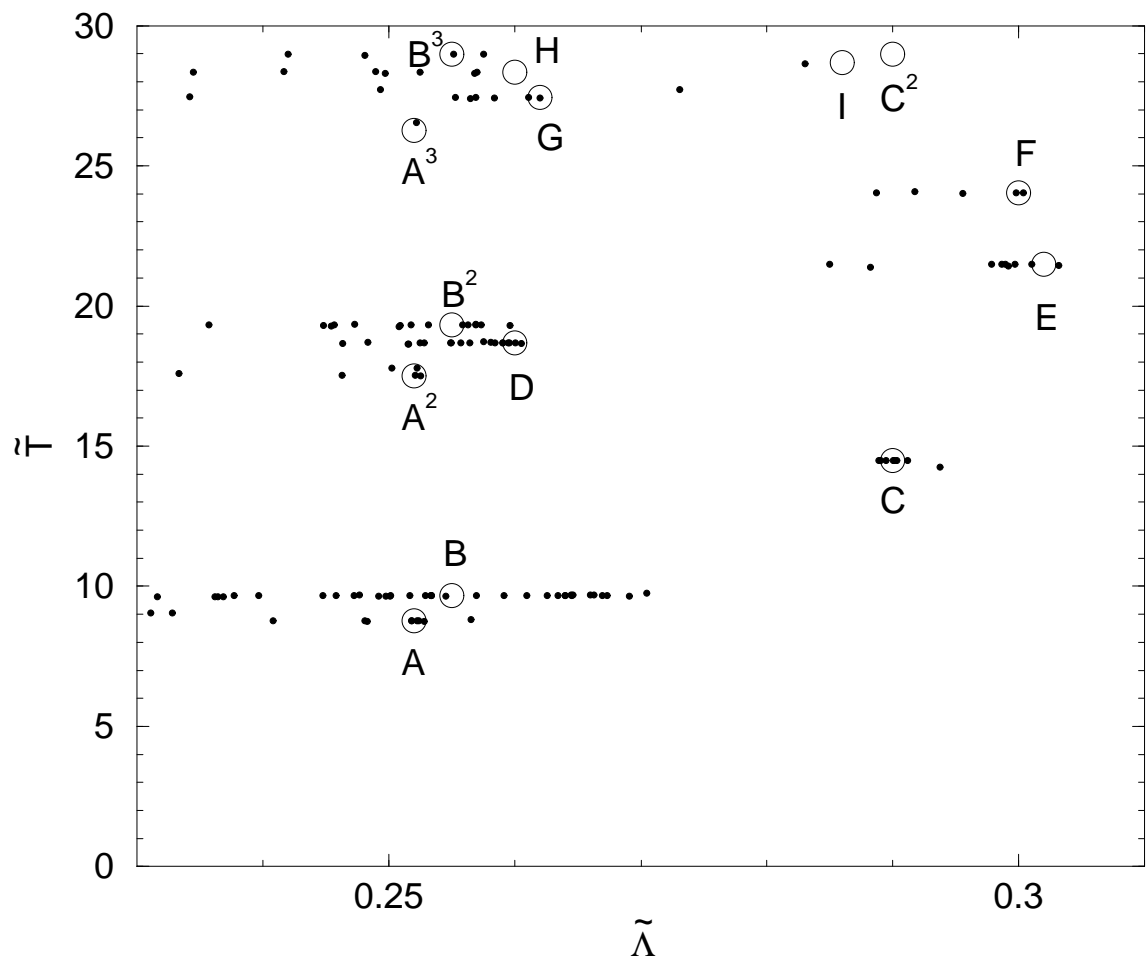}
\caption{Periods and Lyapunov exponents of periodic orbits evaluated by
using the time delay, $\alpha_{\mathrm{out}}$ and $b_{\mathrm{out}}$
scattering functions.
Exact values are marked by open circles and capital letters.
The letter's upper indices display what fraction of the full period
were measured.}
\label{analyzed_062}
\end{center}
\end{figure}
\begin{landscape}
\end{landscape}
\newpage
\begin{table}[ht]
\begin{center}
\begin{tabular}{|c|c|r|r|r|c|}
\hline
orbit&\#&$T_{\mathrm{measured}}$&
$T_{\mathrm{exact}}$&$\Lambda_{\mathrm{measured}}$&
$\Lambda_{\mathrm{exact}}$\\
\hline
A&12& 8.762$\pm$0.014& 8.757&0.245$\pm$0.012&0.252\\
B&41& 9.663$\pm$0.017& 9.661&0.246$\pm$0.028&0.255\\
C&8&14.484$\pm$0.005&14.489&0.290$\pm$0.001&0.290\\
D&19&18.678$\pm$0.016&18.678&0.253$\pm$0.015&0.260\\
E&4&21.467$\pm$0.027&21.485&0.296$\pm$0.009&0.302\\
F&16&24.033$\pm$0.016&24.036&0.283$\pm$0.023&0.300\\
G&7&27.450$\pm$0.021&27.439&0.254$\pm$0.012&0.262\\
H&6&28.345$\pm$0.021&28.333&0.243$\pm$0.015&0.260\\
I&0&&28.688&&0.286\\
\hline
\end{tabular}
\end{center}
\caption{Measured Periods and eigenvalues of periodic orbits.
Sample region was $E=0.062$, $\alpha_{\mathrm{in}}=0$,
$b_{\mathrm{in}}\in [1.3049,1.3065]$. RMS errors.}
\label{Result_Table062}
\end{table}

\end{landscape}

\end{document}